\documentclass[useamsfonts]{pasj00}
\begin{document}
\SetRunningHead{Makiya et al.}{{\it Suzaku} Observation of A1555 and A1558}
\Received{2000/12/31}
\Accepted{2001/01/01}

\title{{\it Suzaku} Observation of Abell 1555 and Abell 1558: Searching for Non-thermal Emission from Large Scale Structure Formation }

\author{Ryu \textsc{Makiya}, Tomonori \textsc{Totani}}
\affil{
Dept. of Astronomy, KyotoUniversity, Kitashirakawa-Oiwake-cho, Sakyo-ku, Kyoto 606-8502}
\email{makiya@kusastro.kyoto-u.ac.jp, totani@kusastro.kyoto-u.ac.jp}

\author{Kazuhiro \textsc{Nakazawa}}
\affil{Department of Physics, The University of Tokyo, 7-3-1Hongo, Bunkyo-ku, Tokyo 113-0033}\email{nakazawa@amalthea.phys.s.u-tokyo.ac.jp}

%


\maketitle

\begin{abstract}
  We report X-ray observations of two galaxy clusters Abell 1555 and
  Abell 1558 with {\it Suzaku}, which are included in a large scale
  filamentary structure and a supercluster, to search for non-thermal
  emission driven by shocks produced in structure formation. These two
  clusters are detected by Suzaku/XIS for the first time in the X-ray
  band of 0.5--7 keV. No significant flux is detected by HXD in the
  energy band of 13--40 keV, and upper limits are reported. From the
  analysis of the XIS data, we find that the spectrum of A1555 is fit by a
  thermal plus power-law model, significantly better than a
  single-temperature pure thermal spectrum.  If this
  power-law component is due to inverse-Compton scattering, the fraction
  of total baryon energy imparted to non-thermal electrons is
  consistent with the typical value inferred from the observation of
  other clusters. However, other scenarios (e.g., under lying AGNs, 
  multi-temperature thermal models) cannot be excluded and further 
  investigation of this system is desired.
  Basic physical properties of A1555 (e.g., total mass) are also
  reported.
\end{abstract}

\section{Introduction}
Galaxy clusters are the largest gravitationally bound structures in
the Universe. It contains $\sim 10^{15} \ M_{\odot }$ hot gas, galaxies,
dark matter, and non-thermal components such as magnetic fields and
cosmic-rays. Although the detailed properties of galaxy clusters are
intrinsically interesting topics, they are also important to
understand the formation and evolution of the large scale structure,
and acceleration mechanisms of cosmic-rays.

According to the cold dark matter scenario, which is the most 
successful theory of the structure formation, structures grow 
up hierarchically from small scale to large scale.
When an object collapses gravitationally and virializes, 
shock waves occurs and a fraction of the baryonic matters in objects 
are accelerated into relativistic energies, if the shock waves are strong enough.
Accelerated electrons scatter the 
Cosmic Microwave Background (CMB) photons up to hard X-ray and 
gamma-ray regions via the Inverse-Compton (IC) mechanism 
(for a recent review of gamma-rays from structure formation, 
see \cite{blasi2007}).

Although the IC emission due to the structure formation has not been
detected at high significance, the existence of relativistic electrons
in several clusters has been revealed by the observations of diffuse
radio emission from galaxy clusters (\cite{1970MNRAS.151....1W};
\cite{Giovannini2000}; \cite{Govoni2004}; \cite{2005AdSpR..36..729F}).
These extended radio emissions are called as``radio halos", which is
located at the cluster center, or``radio relics" at the cluster
periphery. They have steep spectra and a cutoff at a few GHz, implying
that the relativistic electrons account for the radio emission.  The
radio halos or relics are found in the merging clusters, and hence
these electrons are considered to be accelerated in the cosmological
shock waves.

Since cluster X-ray emission is dominated by a thermal component below
$\sim$ 20 keV and expected non-thermal emission has a low flux,
observing the IC emission is a difficult task. Although great effort
has been devoted to search for the non-thermal IC emission, results
are still controversial. For example, Beppo SAX reported the detection
of hard X-ray emission from Coma cluster (\cite{Fusco1999};
\cite{Fusco2004}), but these detections are still a matter of debate
(\cite{Rossetti2004}; \cite{Fusco2007}). Observations of galaxy
clusters in gamma-ray band have also been performed by ground-based
air Cherenkov telescopes and {\it Fermi} (e.g., \cite{hessA496};
\cite{hessCOMA}; \cite{magicPER}; \cite{Fermicluster}), but no
significant detection of non-thermal emission has been achieved, only
resulting in flux upper limits.

Here we report X-ray observations of two galaxy clusters, Abell 1555
and 1558 by {\it Suzaku}, whose redshifts are $z = 0.127$ and 0.116
(Colafrancesco 2002).  
The original motivation of this observation was
to examine a hypothesis that an unidentified EGRET gamma-ray source
3EGJ 1234-1318, which is located at the region of large scale
filamentary structure including these two clusters
(\cite{Einasto2001}; \cite{KT02}; see fig.\ref{fig:matched}), is IC
emission from electrons accelerated by structure formation
shocks. Recently, 3EGJ 1234-1318 was detected by the new generation
gamma-ray telescope {\it Fermi} as 1FGL J1231.1$-$1410, but its
location has changed outside of the region of our Suzaku observation,
and furthermore, its origin has been determined to be a gamma-ray
pulsar (Ransom et al. 2011; Maeda et al. 2011).

However, the region around A1555 and A1558 is still interesting as a
general study for a supercluster region searching for non-thermal
emission clusters.  There have been no targeted deep X-ray
observations for these clusters. We will present the result of our
search for non-thermal nature of X-ray emission, as well as a study of
general physical properties of these two clusters.  The Hard X-ray
Detector (HXD; \cite{hxd2}; \cite{hxd1}) onboard {\it Suzaku}
(\cite{Mitsuda2007}) is characterized by its low detector background
and wide field of view for hard X-rays at 10--50 keV (HXD-PIN) and 
50--600 keV(HXD-GSO). {\it Suzaku} X-ray CCD cameras (XISs; \cite{xis}) 
has a large effective area and a stable and low detector background 
in 0.5--10 keV. These detectors are suitable for observing diffuse 
emission from galaxy clusters.

In section \ref{sec:Obs} we present observation logs and data
reduction method, and section \ref{sec:Results} is devoted for
describing the result of data analysis. In section
\ref{sec:Discussion} we discuss the non-thermal nature of this
supercluster. We will summarize our results in section
\ref{sec:Summary}. Throughout this paper, the cosmological parameters
of $\Omega_{0}=0.3$, $\Omega_{\Lambda}=0.7$, $h = H_{0}/ (100 \ {\rm
  Mpc^{-1}~km~s^{-1}}) =0.7$, and the baryon density parameter
$\Omega_{B} = 0.015h^{-2}$ are assumed.  Unless otherwise noted, all
errors are at the 90\% confidence level.

\section{Observation and Data reduction}
\label{sec:Obs}

\begin{table*}
  \begin{center}
  \caption{Observation log. \label{tb:log}}
   \begin{tabular}{lcccc}
    \hline
    \multicolumn{1}{c}{Obs. region} & Obs. ID & Start (UT) & End (UT) &  Exp. (ks)\footnotemark[$*$] \\
    \hline
	\hline
    Region 1 & 801032010 & Dec 12, 2006 22:16 & Dec 13, 2006 18:38 & 25.0/18.7\\
    Region 2 & 801031010 & Dec 12, 2006 08:56 & Dec 12, 2006 22:15 & 10.7/26.1 \\
	Region 3 & 802004010 & Dec 11, 2007 01:27 & Dec 11, 2007 20:30 & 23.2/26.3\\
    Region 4 & 802005010 & Dec 11, 2007 20:31 & Dec 12, 2007 11:10 & 13.8/20.0\\
   \hline
   \multicolumn{4}{@{}l@{}}{\hbox to 0pt{\parbox{180mm}{\footnotesize
       \par\noindent
       \footnotemark[$*$] The XIS/HXD Exposure time after screening
     }\hss}}
   \end{tabular}
 \end{center}
\end{table*}

\begin{figure*}
   \begin{center}
	  \FigureFile(90mm,36mm){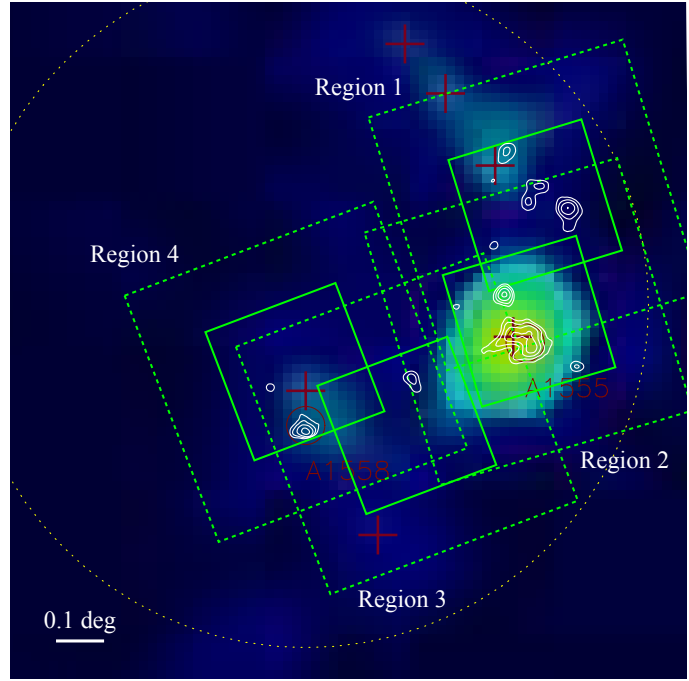}
   \end{center}
   \caption{The cluster ``richness'' map for our observation 
   region overlaid (blue) by X-ray photon count contour (white solid line).
   We use the blue passband data of the APM galaxy catalog to make the ``richness'' map.
   See Kawasaki and Totani (2002) for detail. X-ray photon count contour is obtained by XIS-3
   in the energy range of 0.5 -- 7.0 keV.
   The contour levels are 2, 2.4, 2.8, 3.4, 4.0 (in $10^{-7}$ ${\rm cts\;s^{-1}\;pix^{-1}}$).
   Solid and dotted green boxes are XIS and HXD FoVs, respectively. Plus signs are galaxy cluster detected 
   by Kawasaki and Totani (2002). Dotted yellow circle represents the error circle of 3EG J1234-1318
   (The location of this source has been changed by Fermi observation; see text).
   }
\label{fig:matched}
\end{figure*}

\begin{figure*}
   \begin{center}
      \FigureFile(90mm,36mm){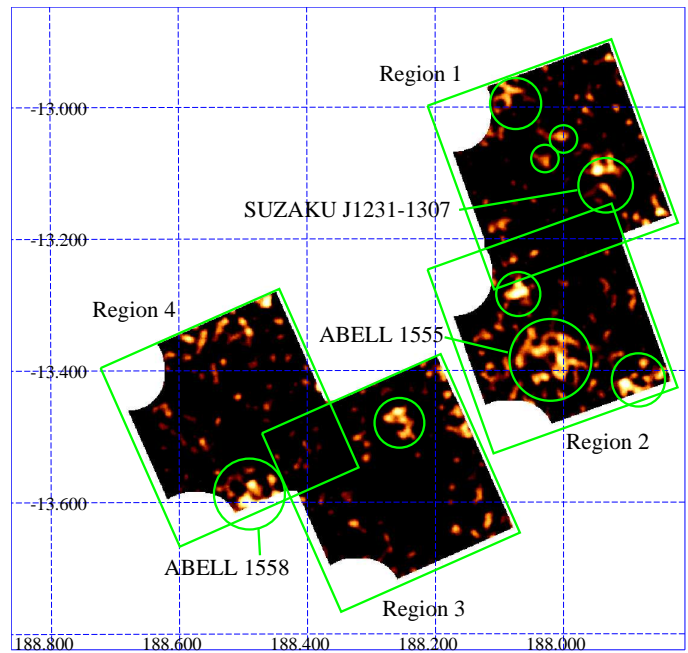}
   \end{center}
   \caption{XIS-3 0.5 keV -- 8.0 keV mosaic image. Green boxes are
     XIS FoV. Images were corrected for exposure time and vignetting
     effect after subtracting NXB, and smoothed by a Gaussian kernel
     with $\sigma=0.5$ arcmin. Small green circles show the detected
     sources.}
\label{fig:xisimage}
\end{figure*}

The observations were carried out on 2006 December 12-13 and 2007
December 11-12 for four pointings. Observation pointings are located 
along the filamentary structure and covers two galaxy clusters, A1555 
and A1558. Hereafter we call these four observation areas as
region 1, 2, 3, and 4, respectively (see figure \ref{fig:matched} and
figure \ref{fig:xisimage}).  Details of the observation dates and
exposure times are summarized in table \ref{tb:log}.  In latter two
observations, the trouble in onboard data processing of XIS-0 occurred
and no usable data was output for the half area, and therefore we do not
use XIS-0 data in this study.

The data for region 1 and 2 were processed with the {\it Suzaku}
pipeline processing of version 2.0.6.13, and region 3 and 4 were
processed with the version 2.2.8.20. We used the calibration data
files of 20081009, and HEASOFT version of 6.6.3. XISs data obtained
with 3$\times$3 and 5$\times$5 edit modes for each observation and we
combined them into one and applied standard filters as follows.
Events with a GRADE of 0, 2, 3, 4, 6, and STATUS $<$ 1024 were
extracted. We selected good time intervals by removing the time that
space craft located at the South Atlantic Anomaly (SAA) within 436
sec, the cut-off rigidity $<$ 6.0 GV, elevation angle from the earth
rim $<$ $5^{\circ }$, and the sun-lit earth rim $<$ $20^{\circ }$.

Major background in {\it Suzaku} observation is Non X-ray Background 
(NXB) and Cosmic X-ray Background (CXB). NXB images and spectra of 
the XISs were created using the ftool ``xisnxbgen'' \citep{tawa2008}. 
\citet{tawa2008} studied the NXB error on a typical observation 
lasting for a few days. At the 90\% confidence level, it is 6.7\% 
and 12.5\% for the sum of two Front Illuminated (FI) XISs (XIS-0,3) 
data and the Back Illuminated (BI) XIS (XIS-1) data, respectively. 
In the following analysis, we utilize this value as a systematic 
error of NXB model. The CXB fluctuation can be modeled as 
$\sigma_{{\rm CXB}}/I_{{\rm CXB}} \propto \Omega_{e}^{-0.5}S_{c}^{0.25}$ (Condon 1974). 
Here, $\Omega_{e}$ is the effective solid angle and $S_{c}$ is the 
upper cut-off flux. From the HEAO-1 A2 results, \citet{Shafer1983} 
derived $\sigma_{{\rm CXB}}/I_{{\rm CXB}}=2.8\%$(1$\sigma$) with 
$\Omega_{e}=15.8\ {\rm deg}^{2}$ and $S_{c}=8 \times 10^{-11}\ 
{\rm erg~s^{-1}~cm^{-2}}$. By scaling this result with XIS parameters, 
we obtain $\Omega_{e}=0.09~{\rm deg^{2}}$ and $S_{c} \sim 4.94 \times 
10^{-14}{\rm erg~s^{-1}~cm^{-2}}$, which is the lowest flux of point 
sources detected in this observation. CXB fluctuation in XISs were 
estimated to be 9.65\% (90\% confidence limit).

The HXD data were also processed in a standard way. We screened the 
data with following criteria: the cut-off rigidity is larger than 
6.0 GV, the elapsed time after the passage of SAA is more than 500 
sec and the time before entering the SAA is more than 180 sec, 
elevation angle from the earth rim $> 5^{\circ }$. We used a public 
NXB model provided by the HXD team for the NXB of the HXD-PIN. 
The version of the model is``METHOD=LCFITDT'' 
or ``tuned'' \citep{Fukazawa2009}.

According to the \citet{Fukazawa2009}, the NXB model reproducibility
of blank sky observations separated into 10 ks exposures give
distribution of 5.7\% at the 90\% confidence level, including the
contribution from the statistical error of typically 3.3\% or larger
and the effect of CXB fluctuation (1.3\% of the total
background). Thus the systematic error of NXB model is calculated to
be 4.5\% at the 90\% confidence level. We use the same model as
\citet{Nakazawa2009} for the CXB of HXD; $N(E)=8.7 \times 10^{-4}
E^{-1.29}\times \exp(-E/40.0)$ in ${\rm
  photons^{-1}~s^{-1}~keV^{-1}~FoV^{-1}}$, where $E$ is the photon
energy in keV. Details of the CXB model estimation are described in
appendix 2 of \citet{Nakazawa2009}. Using the same model of the CXB
fluctuation with the XIS, we obtain the CXB fluctuation in HXD as
18\%, with $\Omega_{e}=0.32~{\rm deg^{2}}$ and conservative upper
cut-off flux of $S_{c} \sim 8 \times 10^{-12}\ {\rm erg~s^{-1}~cm^{-2}}$
in 10--40 keV band.

\section{Results}
\label{sec:Results}
\subsection{HXD results}
\label{sec:HXDresults}

HXD-PIN detected no significant signals above the background in all
regions and we only set upper limits in each observation region using
13 keV -- 40 keV data.  
We converted residual signals (data-(NXB+CXB)) into a energy flux in 13
keV -- 40 keV by assuming a power law emission with photon index
$\Gamma = 2.0$ located at a center of HXD FoV. Results are summarized
in table \ref{tb:hxd}. Systematic errors are the sum of CXB and NXB
model uncertainty as described in section \ref{sec:Obs}.  In all
regions the residual signal is negative. It means that CXB and NXB
models are slightly overestimated, and we derived the flux upper
limits from statistical and systematic errors, assuming physical flux
to be zero.  The comparison of these limits with the expected
non-thermal IC emission is discussed in section \ref{sec:NTfromA1555}.

We also check the HXD-GSO spectrum and confirmed that it is consistent with the 
NXB model.Thus we did not use them.

\begin{table}
  \begin{center}
   \begin{tabular}{lcccc}
    \hline
    \multicolumn{1}{c}{ }  & Flux [${\rm erg~cm^{-2}~s^{-1}}$] & Upper limit  \\
    \hline
  Region 1 & $(-7.4 \pm 2.3 \pm 5.9) \times 10^{-12}$  & $8.2 \times 10^{-12}$\\
  Region 2 & $(-0.7 \pm 2.8 \pm 6.0) \times 10^{-12}$ & $8.8 \times 10^{-12}$\\
  Region 3 & $(-0.1 \pm 2.4 \pm 6.0) \times 10^{-12}$ & $8.4 \times 10^{-12}$\\
  Region 4 & $(-0.2 \pm 2.8 \pm 6.1) \times 10^{-12}$ & $8.9 \times 10^{-12}$\\
   \hline\\
   \end{tabular}
  \end{center}
  \caption{
    Measured flux and upper limits 
    in 13 KeV -- 40 keV from the HXD data. 
    Spectral 
    shape is assumed to be a power law with photon index 2. 
    The former error is statistical, and the latter is systematic, 
    due to CXB and NXB model uncertainty. All errors are at the 90\% confidence level.
	\label{tb:hxd}
}
 \end{table}

\subsection{XIS results}

\begin{table*}
  \caption{The list of sources which are detected in this observation. 
		 See the text for detail. \label{tb:sources}}
  \begin{center}
  \begin{tabular}{lccccc}
  \hline
  \multicolumn{1}{c}{Name} & RA & DEC & Flux \footnotemark[$*$] & Point/Diffuse & Counter part\\
  \hline
  SUZAKU J1232-1259 & 12:32:18 & $-$12:59:41 & $1.4  \pm 0.20  $ & Point   & --\\
  SUZAKU J1232-1302 & 12:32:00 & $-$13:02:55 & $0.62 \pm 0.13 $ & Point   & --\\
  SUZAKU J1232-1304 & 12:32:07 & $-$13:04:42 & $0.49 \pm 0.12 $ & Point   & --\\
  SUZAKU J1231-1307 & 12:31:44 & $-$13:07:08 & -- &  Diffuse & --\\
  SUZAKU J1232-1317 & 12:32:17 & $-$13:17:02 & $1.0  \pm 0.19  $ &  Point   & --\\
  SUZAKU J1231-1323 (A1555) & 12:31:60 & $-$13:23:33 & -- &  Diffuse & --\\
  SUZAKU J1231-1325 & 12:31:30 & $-$13:25:00 & $0.78 \pm 0.28  $ &  Point   & NVSS J123136-132502\\ 
  SUZAKU J1233-1328 & 12:33:02 & $-$13:28:46 & $0.65 \pm 0.24  $ &  Point   & NVSS J123258-132727\\ 
  SUZAKU J1234-1334 (A1558) & 12:34:00 & $-$13:34:32 & -- & Diffuse & --\\
  \hline
   \\
   \multicolumn{6}{@{}l@{}}{\hbox to 0pt{\parbox{180mm}{\footnotesize
       \par\noindent
       \footnotemark[$*$] Energy flux in ${\rm 10^{-13} \  erg\;s^{-1}\;cm^{-2}}$, where errors are at the 90\%
	   					  confidence limit. Flux of three diffuse sources are shown in table \ref{tb:xis}. \\
     }\hss}}
  \end{tabular}
\end{center}
\end{table*}

\subsubsection{Image analysis}
\label{sec:imageanalysis}
In figure \ref{fig:xisimage}, we show the XIS3 mosaic image. Images
were corrected for exposure time and vignetting effect after
subtracting NXB, and smoothed by a Gaussian kernel with $\sigma=0.5$
arcmin.  Diffuse emissions from A1555 and A1558 are clearly
detected, which are the first detection in the X-ray band.  We also
selected seven circular regions centered on significant signals
selected by eyes (represented by green circles in figure
\ref{fig:xisimage}). On the other hand, there are no apparent emission
associated with the filamentary structure (see
figure\ref{fig:matched}).

The non-thermal diffuse emission is expected to be spatially extended.
To examine whether the signals except for the A1555 and A1558 are diffuse 
or not, we compare the spatial expanse of these
signals with the {\it Suzaku} point spread function (PSF). Since the
PSF changes along with the source location on the XISs FoV, we
estimated the PSF at the each source position using the
ftool ``xissim". We simulated the image of point-like sources with the
same exposure time as our observation, and utilize the spatial photon
distribution of simulated point source as the PSF.

We fitted the PSF and photon distribution of observed source with a
Gaussian function. If the dispersions of the PSF and the photon
distribution of an observed source agreed within 1$\sigma$, we regard
this source as a point like source, and otherwise we regard it as a
diffuse source.  As a result, one source is determined to be diffuse
(named ``SUZAKU J1231-1307", see figure \ref{fig:xisimage}) and others
are point-like sources.  We present detailed spectral analysis of this
diffuse source, as well as A1555 and A1558, in next
subsection.

Physical properties of all sources such as positions, photon flux,
point-like or diffuse, and radio counterpart (if exists) are
summarized in table \ref{tb:sources}.  To derive a photon flux of a
source, we assumed a power law model.  Since fitting calculation is
unstable when photon index is free, photon index is fixed at 2. We
selected sources from NLAO VLA Sky Survey (NVSS) catalog (\cite{nvss})
which are located within 0.3 arcmin from the source center as a radio
counter part, where 0.3 arcmin is the typical position determination
accuracy of {\it Suzaku}. We find that SUZAKU J1231-1325 and SUZAKU J1233-1328 
have a radio counterpart, NVSS J123136-132502 and NVSS J123258-132727,
respectively.

\subsubsection{Spectral Analysis: I. Data Reduction}
\label{sec:speana}

We performed spectral fitting of three diffuse sources, A1555,
A1558 and SUZAKU J1231-1307 to search for the non-thermal diffuse
emission.  For the spectral analysis of XIS data, rmf and arf files
were created using {\it Suzaku} ftool ``xisrmfgen" and
``xissimarfgen", respectively.  Two XIS-FI (XIS0, XIS3) spectra and
response files for each sources are summed up, and XIS-FI and XIS-BI
(XIS1) spectra are fitted simultaneously.  Energy band used in the
spectral analysis is 0.5 -- 7.0 keV.

The background spectrum of each FoV were modelled as follows.  First,
we extracted the spectrum from the whole FoV and subtracting the
sources and NXB, and then performed spectral fitting of this data with
the CXB model.  For the CXB model, we used a model including a
power-law component with a fixed photon index ${\rm \Gamma} =
1.4$, a fixed temperature (0.08 keV) thermal component, and another thermal
component whose temperature is a fitting parameter around 0.3 keV (the
so-called ``Galactic components'').  The power-law component
suffers from Galactic absorption, and it was modeled by the {\it wabs}
code.  For the $N_{\rm H\emissiontype{I}}$ column density we use the value of
\citet{Kalberla2005}, which are 3.47, 3.41, 3.41 and $3.43 \times
10^{20} {\rm \ cm^{-2}}$ for the regions 1, 2, 3 and 4, respectively.

Results of spectral fittings to CXB are summarized in table
\ref{tb:CXB}, where the values of fitting model parameters (the three
normalization parameters and a fitting temperature) are shown. 
In each region, our CXB model describes the observed data well (see the
reduced $\chi^2$ in the table), and the 2-10 keV fluxes of the
power-law component are consistent within 90\% statistic and
systematic error with the best-fit value derived using {\it ASCA},
which is $5.67 \pm 0.04 \times 10^{-8} \ {\rm erg~cm^{-2}~s^{-1}~str^{-1}}$ (\cite{Kushino2002}). 
It has been known that solar-wind charge exchange with H\emissiontype{I} in the 
Earth's geocorona accounts for a part of X-ray background below 1keV (e.g., Cravens 2000; Fujimoto et al. 2007).
We do not consider detailed characterization of this component in this paper 
since our CXB model well describes the spectra of background regions.
(Note that this component may be implicitly included in the modeling of the 
Galactic component.)

Then, we simulated the CXB spectrum to produce X-ray counts
using the obtained parameters of the CXB and arf files at each source
position. Finally, we combined the simulated CXB and NXB spectrum, 
and utilized it as a background spectrum in the spectral fitting of sources.

\subsubsection{Spectral Analysis II. Model Fittings}
We fitted the spectra of the three diffuse sources by thermal,
power-law, and thermal plus power-law models.  For the thermal model
we use the {\it apec} code in the XSPEC (v11.3.2).  For the value of
$N_{\rm H\emissiontype{I}}$ column density we also
used the value of \citet{Kalberla2005}, which are $3.49$, $3.41$, and
$3.43 \times 10^{20} {\ \rm cm^{-2}}$ for A1555, A1558 and
SUZAKU J1231-1307, respectively. Since the statistics is limited we could 
not determine the metal abundance by spectral fitting, and therefore we fixed 
it at ${\rm 0.3~Z_{\odot}}$ which is the average value of the clusters in the 
distant cluster catalog compiled by Ota \& Mitsuda (2004).
We adopted the solar abundance table of Anders \& Grevesse (1989).
The redshift was fixed at z = 0.127 for A1555 and z =
0.116 for A1558, respectively \citep{Colafrancesco2002}.  For the
SUZAKU J1231-1307 we fixed the redshift at 0.1, since fitting is
unstable when redshift is set to free due to low statistics. Then the
free parameters are the normalizations of the thermal and power-law
components, temperature, and photon index.  The results of spectral
fittings are summarized in table \ref{tb:xis}, and the spectra are
shown in figure \ref{fig:specreg4}, \ref{fig:speca1555},
\ref{fig:speca1558}, respectively.

As a result, all the three models are consistent with the data of all
the three sources in terms of the value of $\chi^2$, except the
power-law model against A1558 ($5 \%$ probability of getting such a
large $\chi^2$).  However, in the case of A1555 the fit significantly
improves when we fit by the thermal plus power-law model
($\chi^2$/d.o.f = 27.8/44), compared with the pure thermal model
(42.4/46).  The F-test probability is $9.3 \times 10^{-5}$, indicating
that the former is significantly favored than the latter.  We examined
the sensitivity of this result to the CXB fluctuation, by artificially
increasing the CXB flux by 14.8\%, which is 1$\sigma$ of the
fluctuation (see section \ref{sec:Obs} for detail).  The resultant F-test
probability only slightly increases to $4.2 \times 10^{-4}$.
We will discuss implications of those results in \S\ref{sec:NTfromA1555}.

In contrast, the spectrum of A1558 was not well fitted by the
power law model, and $\chi^2$ value did not improve when we fit by the
thermal plus power-law model, compared with the pure thermal model.
Therefore A1558 seems to be a well relaxed cluster.

The spectrum of unknown diffuse source SUZAKU J1231-1307 is well
fitted with all models. We will discuss this object later (see section \ref{sec:NTfromA1555}).

  \begin{table*}
    \begin{center}
  \caption{The best fit parameters of the CXB model for background spectrum. All errors are given 
  		   at the 90\% confidence limit.\label{tb:CXB}}
   \begin{tabular}{lccccc}
    \hline
    \multicolumn{1}{c}{ }  & $F_{\rm PL}$ \footnotemark[$*$] & $N_{\rm cool}$ \footnotemark[$\dagger$] & $N_{\rm hot}$ \footnotemark[$\dagger$] & $kT_{\rm hot}$ \footnotemark[$\|$] & $\chi^{2}$/d.o.f \\
    \hline
  Region 1 & $5.3^{+0.3}_{-0.3}$ & $1.2^{+0.4}_{-0.3}$ & $4.7^{+1.7}_{-1.1}$ & $0.26^{+0.03}_{-0.03}$ & 110.1/96\\
  Region 2 & $5.8^{+0.3}_{-0.3}$ & $1.3^{+0.3}_{-0.3}$ & $7.4^{+1.6}_{-1.4}$ & $0.29^{+0.03}_{-0.03}$ & 93.6/121 \\
  Region 3 & $5.2^{+0.3}_{-0.3}$ & $1.4^{+0.3}_{-0.4}$ & $6.3^{+1.6}_{-1.1}$ & $0.28^{+0.03}_{-0.03}$ & 137.7/116 \\
  Region 4 & $5.1^{+0.4}_{-0.4}$ & $1.5^{+0.4}_{-0.4}$ & $5.4^{+1.6}_{-1.4}$ & $0.29^{+0.05}_{-0.04}$ & 120.5/105 \\
   \hline\\
   \multicolumn{6}{@{}l@{}}{\hbox to 0pt{\parbox{100mm}{\footnotesize
       \par\noindent
       \footnotemark[$*$] Surface brightness of power law component in 2.0 -- 10.0 keV (in $10^{-8}\ {\rm erg\;cm^{-2}\;s^{-1}\;str^{-1}}$). 
       \par\noindent
       \footnotemark[$\dagger$] {Normalization in the thermal model, for the cooler and hotter galactic thermal components (in $10^{-2} \ {\rm cm^{-5}}$ and $10^{-4} \ {\rm cm^{-5}}$, respectively).}
       \par\noindent
       \footnotemark[$\|$] Temperature of the hotter galactic thermal component (in keV).
     }\hss}}	 
   \end{tabular}
 \end{center}
\end{table*}

\begin{table*}
  \caption{The best fit parameters of the thermal, power-law (PL), 
           and thermal plus PL model for the three diffuse sources: 
		   SUZAKU J1231-1307, A1555 and A1558. All errors are at the 90\% confidence limit.
           \label{tb:xis}}
  \begin{center}
   \begin{tabular}{lcccccc}
    \hline
	\hline
    \multicolumn{1}{c}{Object}  & Model & {$N_{th}$\footnotemark[$*$]} & {$kT$ \footnotemark[$\dagger$]} &  {$F_{\rm PL}$ \footnotemark[$\ddagger$]} &  {$\Gamma$ \footnotemark[$\S$]}  & {$\chi^{2}$/d.o.f}\\
 \hline
    SUZAKU J1231-1307& Thermal & $0.9^{+0.2}_{-0.2}$ & $3.7^{+2.3}_{-1.1}$ & -- & -- & 15.3/23\\
            & PL    & -- & -- & $6.5^{+2.3}_{-1.9}$ & $1.9^{+0.26}_{-0.24}$ & 18.5/23 \\
	        & Thermal+PL & $1.0^{+0.4}_{-0.8}$ & $6.3^{+5.6}_{-5.6}$ & $1.6^{+7.6}_{-1.6}$ & {\rm 2.0 (fixed)} & 16.9/22\\
 \hline
  A1555 & Thermal & $3.0^{+0.3}_{-0.3}$ & $2.0^{+0.54}_{-0.33}$ & -- & -- & 42.4/46 \\
             & PL    & -- & -- & $10.0^{+2.8}_{-2.4}$ & $2.5^{+0.22}_{-0.20}$ & 33.0/46 \\
			 & Thermal+PL & $0.8^{+0.9}_{-0.6}$ & $1.1^{+0.39}_{-0.22}$ & $10.0^{+3.6}_{-3.4}$ & $2.2^{+0.33}_{-0.46}$ & 27.8/44 \\
 \hline
  A1558 & Thermal & $1.6^{+0.4}_{-0.5}$ & $1.4^{+0.31}_{-0.27}$ & -- & -- & 37.3/29\\
             & PL  & -- & -- & $4.8^{+2.7}_{-2.0}$ & $2.7^{+0.43}_{-0.37}$ & 46.7/29\\
             & Thermal+PL & $1.4^{+0.6}_{-0.4}$ & $1.3^{+0.42}_{-0.27}$ & $1.0^{+4.3}_{-1.0}$ & $2.0$ (fixed) & 37.1/27\\
   \hline
   \\
   \multicolumn{7}{@{}l@{}}{\hbox to 0pt{\parbox{180mm}{\footnotesize
       \par\noindent
       \footnotemark[$*$] Normalization of the thermal model (in $10^{-4}\;{\rm cm^{-5}}$). 
       \par\noindent
       \footnotemark[$\dagger$] Temperature (in keV).
       \par\noindent
       \footnotemark[$\ddagger$] Flux of power law component. 
   (in $10^{-14}\;{\rm erg~cm^{-2}~s^{-1}}$ over the energy range 2.0 -- 10.0 keV)
       \par\noindent
       \footnotemark[$\S$] Photon index of power law component
       \par\noindent
     }\hss}}
   \end{tabular}
 \end{center}
 \end{table*}

\begin{figure*}
   \begin{center}
      \FigureFile(53mm,50mm){figure3a.ps}
	  \FigureFile(53mm,50mm){figure3b.ps}
	  \FigureFile(53mm,50mm){figure3c.ps}
   \end{center}
   \caption{Spectrum of diffuse source, SUZAKU J1231-1307 with the
     power law model, thermal model, and thermal $+$ power law model
     (left to right).  Data points of XIS-FI are represented by black,
     while XIS-BI by red. Solid line represents best-fit model.
 See Table \ref{tb:xis} for fitted model parameters. 

}
	\label{fig:specreg4}

   \begin{center}
      \FigureFile(53mm,50mm){figure4a.ps}
	  \FigureFile(53mm,50mm){figure4b.ps}
	  \FigureFile(53mm,50mm){figure4c.ps}
   \end{center}
   \caption{The same as figure \ref{fig:specreg4}, but for Abell1555.}\label{fig:speca1555}

   \begin{center}
      \FigureFile(53mm,50mm){figure5a.ps}
	  \FigureFile(53mm,50mm){figure5b.ps}
	  \FigureFile(53mm,50mm){figure5c.ps}
   \end{center}
   \caption{The same as figure \ref{fig:specreg4}, but for Abell 1558.}\label{fig:speca1558}
\end{figure*}

\subsection{Total Mass estimation of A1555}
\label{sec:mass}

Here we present a detailed estimation of the total mass of A1555, $M_{\rm tot}$,
because this source has been detected with the highest significance and the 
whole cluster is included in the field of view.

In the isothermal $\beta$--model(Cavaliere \& Fusco-Femiano 1976), 
the radial profile of X-ray surface brightness is written as
\begin{equation}
\label{eq:sx}
S_{x}(b) = S_{0} \left [ 1+ \left(\frac{b}{r_{c}} \right)^{2} \right] ^{-3 \beta + 1/2},
\end{equation}
where $b$, $S_{0}$, $r_{c}$ and $\beta$ are the projected radius, central surface 
brightness, core-radius and the outer slope of density profile, respectively.

We simulated the thermal emission from A1555 which follows $\beta$ model profile
using the ftool ``xissim'', and compare it with the observation to estimate the value of
$\beta$ and $r_{c}$. As a spectral model, we used a single temperature thermal model 
with the parameters listed in table \ref{tb:xis}.

\begin{figure}
   \begin{center}
	  \FigureFile(80mm,50mm){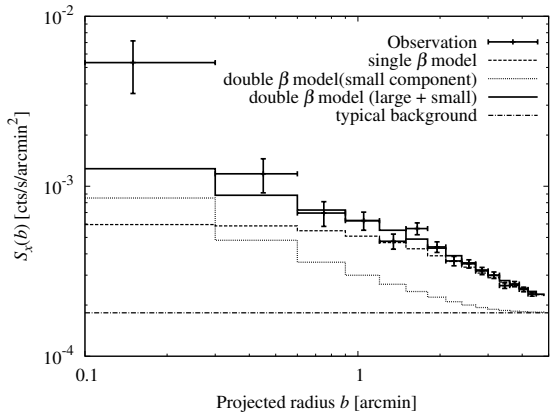}
   \end{center}
   \caption{The surface brightness profile of Abell 1555. The crosses are XIS-BI data.
   		The dashed line represents best-fit single $\beta$ model. The solid line denotes
		double $\beta$ model while the dotted line is small scale component double-$\beta$ model. 
		The typical background level has shown by dot-dashed line.}
   \label{fig:betafit}
\end{figure}

In figure \ref{fig:betafit}, we show the observed surface brightness profile of A1555 with 
the results of simulations.  First, we simulated with $\beta = 0.5$ and $r_{c} = 0.16$ Mpc
(shown as ``single $\beta$ model'' in figure \ref{fig:betafit}).
This value of $r_{c}$ is an average of the clusters located at z = 0.1 -- 0.15 in the distant 
cluster catalog (\cite{Ota2004}). The dispersion of $r_{c}$ is about factor two, and we confirmed 
that this uncertainty of  $r_{c}$ does not significantly affect the estimation of the total cluster mass 
(about factor 0.5). The value of $\beta$ is selected to fit the observation. This value is in agreement 
with the average of the distant clusters (\cite{Ota2004}).
It can be seen that the simulation well reproduces the observed surface brightness profile at the region except
the central two data points.

The central excess is often represented by an additional small scale $\beta$-model 
component (double-$\beta$ model; Jones \& Forman 1984). 
We therefore performed a simulation by adding a small scale component of $r_{c}$ = 
10 kpc/$h$ and $\beta$ = 0.7, in addition to the single $\beta$ model used above, as a
test of the double $\beta$ model.
Those values of $r_{c}$ and $\beta$ are averaged values of the small scale components in 
Mulchaey and Zabludoff (1998). The simulation result is shown in figure \ref{fig:betafit}. 
The double $\beta$ model gives a better fit, but is still discrepant with the innermost data by about
two sigma. This may be just a statistical fluctuation, or may indicate a contribution from a AGN.
Since it is difficult to further discuss the origin of this excess due to the low 
statistics of our data and limitation of spatial resolution of {\it Suzaku},
we conservatively consider a systematic error of a factor of about 2 for $\beta$,
from the dispersion of $\beta$ in the distant cluster sample of Ota \& Mitsuda (2004).

Under the isothermal $\beta$ model and also assuming that the cluster gas is in 
hydrostatic equilibrium, we can estimate the total cluster mass from the value of $r_{\rm c}$, $\beta$, 
and gas temperature (Cavaliere \& Fusco-Femiano 1976).
Using the single $\beta$ model with $\beta = 0.5$ and $r_{c} = 0.16$ Mpc, $M_{\rm tot}
(<r_{500})$ is estimated to be $7.43^{+4.1}_{-2.2} { }^{+14.5}_{-5.0} \times 10^{13} M_{\odot }$ and $r_{500} = 
0.69^{+0.11}_{-0.10} { }^{+0.3}_{-0.2}$ Mpc, where the $r_{500}$ is a radius where the ratio of total cluster mass 
density to critical density became 500. In the above mass estimation we used the value of gas temperature as 
$2.0^{+0.54}_{-0.33}$ keV which is derived from the single thermal component fit (see table \ref{tb:xis}). If we 
used the temperature derived 
from the thermal+PL fit, $1.1^{+0.39}_{-0.22}$ keV, and the same value of $\beta$ and $r_{c}$ assuming that 
thermal and non-thermal component have the same radial profile, $M_{\rm tot}(<r_{500})$ becomes $2.8^{+1.7}
_{-0.8} { }^{+5.8}_{-2.0} \times 10^{13} M_{\odot }$. The former errors in $M_{\rm tot}$ and $r_{500}$ are 
statistical, simply estimated from statistical errors of $kT$ and $\beta$, while the latter ones are systematic, 
estimated from the systematic error of $\beta$. 
The value of total cluster mass and temperature are consistent with the 
known correlation between the total cluster mass and temperature (e.g., \cite{Ota2004}).

If we adopt the double $\beta$ model, the total cluster mass becomes $7.43 \times 10^{13} 
M_{\odot }$ (with thermal model) or $2.8 \times 10^{13} M_{\odot }$ (with thermal + PL model). 
The difference between the mass estimated with single $\beta$ model and double 
$\beta$ model is less than 1$\%$.

In the above mass estimation, we assumed that A1555 is isothermal.
We checked the effect of this assumption on the mass estimation 
by comparing the total cluster mass of several clusters estimated by Fukazawa et al. (2004)
with the total cluster mass estimated by Vikhlinin et al. (2006), where the former assumed
isothermal while the latter did not. 
These two differ by $\sim20\%$, indicating a similar level of systematic errors for our
mass estimates.

We also derive a luminosity of A1555 in 0.2 -- 10.0 keV band 
from spectral fitting with thermal model. It is estimated to 
be $8.8 \pm 1.6 \times 10^{43}\ {\rm erg\;s^{-1}}$.

\section{Discussion}
\label{sec:Discussion}
\subsection{On the Evidence for Non-thermal Emission from A1555}
\label{sec:NTfromA1555}
Here we discuss the origin of the non-thermal emission implied for
A1555. First we discuss the possibility that it is due to the IC
scattering of CMB photons by accelerated electrons.  From the $\Gamma
= 2.2$ power-law component of the thermal+PL fit for A1555, we
estimate the total energy of non-thermal electrons to be
$3.1^{+1.1}_{-1.1} \times 10^{59}$ erg assuming $z = 0.127$, in the
electron energy range 0.4--1.5 GeV corresponding to the X-ray energy
range of 0.5--7.0 keV. The total thermal energy of baryon gas when 
the structure formation occur is given by $E_{\rm baryon} \sim (3/4)
(\Omega_{B}/\Omega_{0})M_{\rm tot}V_{c}^{2}$, where $V_{c}$ is circular 
velocity of the halo, and which can be calculated from spherical collapse model 
(Peebles 1980: Kitayama and Suto 1996) which is widely used in study of 
structure formation. If we use the total mass of A1555 estimated with 
$kT = 1.1$ keV from the thermal+PL fit (see section \ref{sec:mass}) 
$E_{\rm baryon}$ becomes $7.0^{+8.5}_{-3.1} {}^{+15.0}_{-4.8} \times 10^{60} \ {\rm erg}$,
where the former error is statistic while the latter is systematic which are estimated from
the uncertainty of total mass (see section \ref{sec:mass} for detail).
Therefore the fraction of non-thermal electron energy becomes $\xi_{e} \sim 0.044$, 
which is consistent with the typical value discussed in the literature 
(e.g., Fusco-Femiano et al. 1999, but see also Rossetti \& Molendi 2004).

From the flux of non-thermal PL component and the strength of magnetic field, $B$, we can 
estimate the flux of possible synchrotron emission due to the same electron population 
which emits the non-thermal IC emission.
As already noted in section \ref{sec:imageanalysis}, A1555 has no radio counter part in NVSS catalog.
Flux limit of NVSS is 2.5 mJy/beam and beam size is 45".
Assuming that the synchrotron emission has the same
 spectral index with the PL component, $\Gamma = 2.2$, and extended about 1 Mpc,
 the expected brightness at 1.4 GHz is estimated to be 2.2 mJy/beam with $B$ = 1 $\mu$G.
The typical value of $B$ is 0.1 -- 1 $\mu$G, and therefore the scenario of non-thermal 
IC emission is not inconsistent with the radio observation.

We also discuss the detectability of the non-thermal IC emission with
{\it Fermi} Large Area Telescope (Atwood et al. 2009).  Theoretically
it is reasonable to expect that the IC emission extends to the GeV energy
band with a photon index of $\Gamma \sim $ 2 (e.g., Totani \& Kitayama
2000).  Extrapolating from the observed X-ray flux, the expected flux
in the energy higher than 100 MeV became $F_{\rm > 100 MeV} = 3.9 \times 10^{-10}$
and $4.4 \times 10^{-11}\ {\rm photons\;cm^{-2}\;s^{-1}}$ for $\Gamma =
2.0$ and 2.2, respectively.  These fluxes are about one order of
magnitude lower than the sensitivity of {\it Fermi}, and 
the {\it Fermi} observation does not give a strong constraint.

It should also be noted that the flux of power law component in 13.0 -- 40.0 keV 
is consistent with the flux upper limit obtained by HXD analysis.

It has been known that relaxed clusters do have a cool core and a radially 
decreasing temperature at large radii, and merging clusters have shock heated 
structures. To approximate these effects on our results we fitted
the two temperature thermal model to the spectrum of A1555.
The fitting results are summarized in table \ref{tb:a1555-2kT}.  
The fit also significantly improves comparing with the single temperature fit.  
The thermal+PL model has a smaller $\chi^2$ value than the two temperature
model, but the likelihood ratio is just a factor of about four.
Therefore the two temperature model provides another good description
of the data, as an alternative to the thermal+PL model.  The two
temperatures are determined to be $1.0^{+0.34}_{-0.28}$ keV and
$3.6^{+4.5}_{-2.1}$ keV, which are similar to the result of two
temperature analysis of the Virgo and Centaurus cluster (Fukazawa et
al. 2000).

We also check a possibility that the origin of the power-law component
is the background (or foreground) AGNs. We found two radio point
sources at the outer region of A1555, NVSS J1232-1323 and NVSS 1232-1321 
(Condon et al. 1998). Their flux density at 1.4 GHz are $7.0 \pm 0.5$ 
(NVSS J1232-1323) and $4.0 \pm 0.6$ (NVSS J1232-1321) mJy. If we extrapolate 
their flux to the X-ray regime using the spectral energy distribution model 
of AGNs constructed by Elvis et al. (1994), the expected flux become $0.8
\times 10^{-14}$ -- $1.0\times 10^{-12}$ ${\rm erg\;s^{-1}\;cm^{-2}}$ in 
2.0--10.0 keV. The estimated flux of PL component in the spectrum of A1555 
is $1.0\times 10^{-13}$ ${\rm erg\;s^{-1}\;cm^{-2}}$ in 2.0--10.0 keV, and therefore 
the X-ray flux extrapolated from the flux of radio sources are consistent with 
the flux of PL component within the uncertainty. In conclusion we can not 
reject the scenario that a AGN explains the non-thermal component in 
the spectrum of A1555.

\subsection{On SUZAKU J1231-1307}
\label{sec:J1231}
We also detected interesting diffuse object, SUZAKU J1231-1307, which locates along the 
large scale filamentary structure. The spectrum of SUZAKU J1231-1307 is well fitted by 
power-law, thermal, and thermal + PL model with similar $\chi^{2}$ value.
If SUZAKU J1231-1307 is a galaxy cluster, the temperature is 3.7 keV and the total cluster mass is estimated 
to be $2.6^{+2.7}_{-1.0} \times 10^{14} M_{\odot}$. Here we used the correlation between the total mass
and temperature obtained by Vikhlinin et al. (2009).
If the PL component really exists, the total energy of non-thermal electron becomes 
$5.0 ^{+23.6}_{-5.0} \times 10^{58}$ erg, which is about 15\% of A1555.
Further observation is required to determine the origin of this mysterious object.

\subsection{IC emission upper-limit associated with the large-scale structure as a whole}
\label{sec:ICfromLSS}
If the non-thermal IC emission is emitted from the whole region of the 
large scale filamentary structure and extended more than XIS FoV,
the flux of this emission has some contribution to the background flux.
From the error estimation of CXB flux, we derive an upper limit to this component.
As already noted in section \ref{sec:Obs}, the systematic error of CXB flux in one 
XIS FoV is estimated to be 9.65\%. Combining this uncertainty with the statistical error
in CXB estimation (see table \ref{tb:CXB}), total uncertainty of the CXB emission from 
all XIS FoV becomes 5.4\%. From this uncertainty and the average value of CXB surface brightness, 
the upper limit on the surface brightness of non-thermal IC emission is estimated to be 
2.9 $\times 10^{-9}\;{\rm erg~cm^{-2}~s^{-1}~str^{-1}}$.
The total FoV of four XISs is $1.1 \times 10^{-4}$ str, and therefore the flux upper limit 
of non-thermal IC emission becomes 3.2$\times 10^{-13}\;{\rm erg~cm^{-2}~s^{-1}}$. 
This limit is in the same order of magnitude as the flux of power-law component in 
the spectrum of A1555, 1.0$\times 10^{-13}\;{\rm erg~cm^{-2}~s^{-1}}$.

\begin{table*}
  \caption{Best fit parameters of two-temperature thermal model fitted to the A1555.
           All errors are at the 90\% confidence limit. \label{tb:a1555-2kT}}
  \begin{center}
   \begin{tabular}{ccccc}
    \hline
    \multicolumn{1}{c}{$kT_{\rm low}$ \footnotemark[$*$]} & {$N_{\rm low}$ \footnotemark[$\dagger$]} & {$kT_{\rm hi}$ \footnotemark[$*$]} & {$N_{\rm hi}$ \footnotemark[$\dagger$]} & {$\chi^2$/d.o.f}\\
    \hline
		$1.0^{+0.3}_{-0.3}$ & $0.9^{+1.2}_{-0.5}$ & $3.6^{+4.5}_{-2.1}$ & $1.9^{+0.6}_{-1.2}$ & 30.6/44\\
   \hline\\
   \multicolumn{3}{@{}l@{}}{\hbox to 0pt{\parbox{85mm}{\footnotesize
       \footnotemark[$*$] Temperature of cooler (hotter) component (in keV).\\
	   \footnotemark[$\dagger$] Normalization of the {\it apec} code, for cooler and hotter component (in $10^{-4}\ {\rm cm^{-5}}$).
     }\hss}}
   \end{tabular}
 \end{center}
 \end{table*}

\section{Summary}
\label{sec:Summary}

We performed the deepest X-ray observation of the supercluster
region including A1555 and A1558 with {\it Suzaku}. 
This region is interesting since large scale filamentary structures
connected with the two galaxy clusters exist, but only RASS data was
available in the X-ray band.

We have detected the two galaxy clusters A1555 and A1558 for the first
time in the X-ray band.  Six point sources and one diffuse source
(SUZAKU J1231-1307) were also detected. The total mass of A1555 was derived
using the beta model.  We examined the non-thermal nature of SUZAKU
J1231-1307, A1555, and A1558, and there are no evidence for
non-thermal emission for SUZAKU J1231-1307 and A1558. However, we
found that the A1555 spectrum cannot be fit by one-temperature
thermal model.  A thermal plus power-law model significantly improves
the fit, and if this power-law component is the IC emission of CMB
photons, the non-thermal electron energy is estimated to be 2\% of the
thermal energy of the cluster, which is similar to the values
often discussed in the literature for galaxy clusters.

However, the A1555 spectrum can also be fit by a two-temperature
thermal model, and the possibility of underlying AGNs cannot be
excluded.  More detailed and sensitive observational studies are
desired for better understanding of this interesting region.


\bigskip
This work was supported by the Grant-Aid for the Global
COE Program ``The Next Generation of Physics, Spun from
University and Emergenc'' from the Ministry of Education,
Culture, Sports, Science and Technology (MEXT) of Japan.
TT and KN was supported in part by the Grant-in-Aid (22244019) 
and Grant-in-Aid (18104004), respectively, for Scientific
Research from MEXT.

\end{document}